\documentstyle[12pt,equations]{article}
\begin{document}
\newcommand{\Ab}{\bar{A}}
\newcommand{\eb}{\bar{e}}
\newcommand{\et}{\tilde{e}}
\newcommand{\thone}{\theta_1}
\newcommand{\thtwo}{\theta_2}
\newcommand{\tcr}{T_{cr}}
\newcommand{\chit}{\tilde{\chi}}
\newcommand{\phit}{\tilde{\phi}}
\newcommand{\df}{\delta \phi}
\newcommand{\dr}{\delta \rho}
\newcommand{\dkl}{\delta \kappa_{\Lambda}}
\newcommand{\dkg}{\delta \kappa_{G}}
\newcommand{\dkcr}{\delta \kappa_{cr}}
\newcommand{\dxg}{\delta x_{G}}
\newcommand{\dx}{\delta x}
\newcommand{\lx}{\lambda}
\newcommand{\Lx}{\Lambda}
\newcommand{\ex}{\epsilon}
\newcommand{\ebs}{\bar{e}^2}
\newcommand{\ks}{k_s}
\newcommand{\gb}{\bar{g}}
\newcommand{\lb}{{\bar{\lambda}}}
\newcommand{\lbz}{\bar{\lambda}_0}
\newcommand{\lt}{\tilde{\lambda}}
\newcommand{\lr}{{\lambda}_R}
\newcommand{\lrt}{{\lambda}_R(T)}
\newcommand{\lbr}{{\bar{\lambda}}_R}
\newcommand{\lk}{{\lambda}(k)}
\newcommand{\lbk}{{\bar{\lambda}}(k)}
\newcommand{\lbkt}{{\bar{\lambda}}(k,T)}
\newcommand{\ltk}{\tilde{\lambda}(k)}
\newcommand{\mx}{{m}^2}
\newcommand{\mxk}{{m}^2(k)}
\newcommand{\mxkt}{{m}^2(k,T)}
\newcommand{\mb}{{\bar{m}}^2}
\newcommand{\mbb}{{\bar{m}}_2^2}
\newcommand{\mbu}{{\bar{\mu}}^2}
\newcommand{\mt}{\tilde{m}^2}
\newcommand{\mr}{{m}^2_R}
\newcommand{\mrt}{{m}^2_R(T)}
\newcommand{\mk}{{m}^2(k)}
\newcommand{\rht}{\tilde{\rho}}
\newcommand{\rhz}{\rho_{0R}}
\newcommand{\rhzt}{\rho_0(T)}
\newcommand{\rhztil}{\tilde{\rho}_0}
\newcommand{\rhzk}{\rho_0(k)}
\newcommand{\rhzkt}{\rho_0(k,T)}
\newcommand{\kx}{\kappa}
\newcommand{\kt}{\tilde{\kappa}}
\newcommand{\kk}{\kappa(k)}
\newcommand{\ktk}{\tilde{\kappa}(k)}
\newcommand{\Gammat}{\tilde{\Gamma}}
\newcommand{\Gammak}{\Gamma_k}
\newcommand{\wt}{\tilde{w}}
\newcommand{\be}{\begin{equation}}
\newcommand{\ee}{\end{equation}}
\newcommand{\een}{\end{subequations}}
\newcommand{\ben}{\begin{subequations}}
\newcommand{\beq}{\begin{eqalignno}}
\newcommand{\eeq}{\end{eqalignno}}
\def \lta {\mathrel{\vcenter
     {\hbox{$<$}\nointerlineskip\hbox{$\sim$}}}}
\def \gta {\mathrel{\vcenter
     {\hbox{$>$}\nointerlineskip\hbox{$\sim$}}}}

\pagestyle{empty}
\noindent
\begin{flushright}
CERN-TH/96-331 \\
\end{flushright} 
\vspace{3cm}
\begin{center}
{{ \large  \bf
Comparison of Renormalization-Group and Lattice Studies \\
\vspace{0.2cm} 
of the Electroweak Phase Transition 
}}\\
\vspace{10mm}
N. Tetradis 
\vspace {0.5cm}
 \\
{\em 
CERN, Theory Division, \\
CH-1211, Geneva 23, Switzerland
\footnote{E-mail: tetradis@mail.cern.ch.} 
} 

\end{center}

\setlength{\baselineskip}{20pt}
\setlength{\textwidth}{13cm}
 
\vspace{3.cm}
\begin{abstract}
{
We compare the results of renormalization-group 
and lattice studies for the
properties of the electroweak phase transition.
This comparison reveals the 
mechanisms that underlie the phenomenology of the phase
transition.
}
\end{abstract}
\vspace{3cm}
\noindent
CERN-TH/96-331 \\
November 1996
\clearpage
\setlength{\baselineskip}{15pt}
\setlength{\textwidth}{16cm}
\pagestyle{plain}
\setcounter{page}{1}

\newpage

The most important consequence of 
the restoration of the electroweak symmetry at high temperatures
\cite{original} is related to the possibility that the  
baryon asymmetry of the Universe has been created
during the electroweak phase transition \cite{baryon}. 
The precise determination of the generated baryon number
depends very sensitively on the details of the
phase transition. This fact has instigated numerous 
studies of its characteristics during the last few years. 
(For recent reviews see ref. \cite{review}.)
The perturbative approach for the determination of 
the temperature-dependent effective potential 
(from which the properties of the phase transition 
can be inferred)  
has been pursued up to two loops \cite{twoloop}.
However, 
the perturbative expansion breaks 
down near and in the symmetric phase of 
gauge theories, due to the appearance of infrared divergences
\cite{gaugediv}.  
In order to overcome this difficulty, alternative approaches have
been followed. Gap equations (truncated versions of Schwinger-Dyson
equations) \cite{buchgap,owe}
have been employed in order to 
obtain systematic resummations of infinite subclasses
of perturbative contributions.
The $\ex$-expansion \cite{arnold} has also been used in order to
obtain insight into the non-perturbative character of the phase transition. 
The most reliable quantitative results have been obtained through the
lattice approach \cite{fodor1}--\cite{ilg}. However, the underlying 
dynamics that results in a certain physical behaviour is often
obscured by the Monte-Carlo simulations. The analytical approaches
offer a more intuitive understanding. Also, the requirement of 
long computer time for the simulations means that 
the exploration of the full phase diagram for a particular
model is often a formidable task. 

In ref. \cite{me}
a different approach has been followed, based on the
exact renormalization group \cite{wilson}. 
The method of the 
effective average action \cite{averact} has been employed. 
The effective average action
$\Gamma_k$ results from the integration of fluctuations with characteristic
momenta larger than a given scale $k$ (that can be identified with 
the coarse-graining scale of the system).
This scale acts as an effective
infrared regulator and gives control over the regions
in momentum space from which divergences are expected to arise in 
perturbation theory. The dependence of $\Gamma_k$ on $k$ is described
by an exact renormalization-group equation \cite{exact,gauge}.
For large values of $k$ (of the order of the ultraviolet cutoff 
$\Lx$ of the theory) 
the effective average action is equal to the classical action 
(no fluctuations are integrated out), while for $k \rightarrow 0$
it becomes the standard effective action (all fluctuations are 
integrated out). 
As a result, the solution
of the exact renormalization-group equation, with the classical
action as initial condition, gives all the physically relevant information
for the renormalized theory at low scales. The formalism is 
constructed in Euclidean space, which makes the consideration of
temperature effects straightforward, through the imposition of periodic
boundary conditions in the time direction (for bosonic fields) \cite{trans}.  
This approach has been used in the past for the discussion of the
high-temperature phase transitions for the $O(N)$-symmetric scalar theory 
\cite{trans,on}, multi-scalar theories \cite{twoscalar}, and the 
Abelian Higgs model \cite{abelian1,abelian2,me}. 
The high-temperature
phase transition for the $SU(2)$ Higgs model, which
has all the qualitative characteristics of the electroweak phase
transition, has been discussed in refs. \cite{runngauge,bastian,me}.
The phase diagrams of the above models 
(with various fixed and tricritical points)
have been determined in terms of the 
evolution of the potential, which incorporates all the
running couplings of the theory.
Non-universal quantities (critical temperatures)
as well as universal ones (critical exponents and 
amplitudes, the critical equation of state,
crossover curves) have been computed.
A description of the underlying mechanisms
(such as dimensional reduction or radiative symmetry breaking)
has been given. 
The use of the coarse-graining
scale $k$ has provided the necessary framework 
for the proper discussion of 
first-order phase transitions in the context of
Langer's picture of nucleation \cite{me,coarse}.
Overall, 
the method of the effective average action has provided 
a detailed and intuitive picture of the phase transitions in all the above
field theories, without encountering any problem of infrared 
divergences.  
Effective theories of QCD have also been considered in ref. \cite{qcd}.

In this letter we compare the renormalization-group approach 
of ref. \cite{me} with the lattice studies of refs. 
\cite{fodor1}--\cite{laine}. For this, we calculate
the characteristics of the phase transition for the
$SU(2)$ Higgs model, following the procedure 
presented in ref. \cite{me}, for the sets of parameters (gauge
coupling, gauge and Higgs boson masses) used in refs. 
\cite{fodor1}--\cite{laine}. This comparison can be used in order to 
establish the level of accuracy of the renormalization-group approach. 
In spite of the clear and intuitive description of a
large range of physical behaviours through the formalism of the
effective average action,
the quantitative accuracy of the results 
is not easily determined. 
This is due to the approximations that one
has to employ in order to bring the exact renormalization-group
equation for $\Gamma_k$ into a manageable form
that permits its solution.
The largest source of uncertainty is introduced 
by the  assumption for the invariants appearing in  
$\Gamma_k$.
The most commonly employed ansatz is a truncated expression
that keeps the full field and scale dependence of the potential, and 
two-derivative kinetic terms for the fields, 
but neglects higher derivative terms in the effective average action.
An intrinsic estimate of the induced error 
requires a comparison with more extended truncations.
An alternative way of checking the accuracy of the 
results is to compare them with 
the results of other approaches. For example, 
the accuracy of 
the critical exponents for the second-order phase transitions
of the $O(N)$-symmetric scalar theory, 
calculated in refs. \cite{trans,on}, 
has been determined through the comparison with $\epsilon$-expansion
results. This has permitted an estimate of the accuracy
of the result for the critical equation of state. 
For the work of ref. \cite{me}, the comparison with 
the lattice calculations will lead to the understanding
of the mechanisms underlying the phenomenology of the phase transition. 
These mechanisms are obscured by the Monte Carlo simulations, while 
they are clearly observed in the renormalization-group approach.

We give first a brief summary of the approach followed in ref. 
\cite{me} for the study of the $SU(2)$ Higgs model with one doublet. 
An ansatz is employed for the effective average action, which 
includes a general effective average potential $U_k(\rho)$ for the 
scalar field ($\rho=\phi^{\dagger}\phi$), two-derivative kinetic terms
and a gauge-fixing term. The anomalous dimension of the
scalar field is neglected, while the wave-function renormalization 
of the gauge field is incorporated in the running gauge coupling
$e^2(k,\rho)$. 
With the above approximations, the exact renormalization-group equation 
for the $k$ dependence of $\Gamma_k$ can be
translated into evolution equations 
for $U_k(\rho)$ and $e^2(k,\rho)$.
The temperature effects are taken into account through periodic 
boundary conditions in the imaginary-time direction. 
This leads to the generalization of 
the evolution 
equations for the temperature-dependent $U_k(\rho,T)$ and $e^2(k,\rho,T)$.
For $k \gg T$ the temperature 
effects are negligible and the evolution is
typical of a four-dimensional, zero-temperature theory. 
For $k \ll T$ the evolution equations
can be cast in a form typical of a three-dimensional, zero-temperature
theory. This can be achieved 
if one defines effective three-dimensional parameters
by multiplying with appropriate powers of $T$
\be
^3U_k(\rho_3) = \frac{U_k(\rho,T)}{T},~~~~~~~~~~
\rho_3 = \frac{\rho}{T},~~~~~~~~~~
{e}^2_3(k,\rho) = e^2 (k,\rho,T) T,
\label{one} \ee
and their dimensionless versions (rescaled by powers of $k$)
\be
u_k(\rht) = \frac{^3U_k(\rho_3)}{k^3}=\frac{U_k(\rho,T)}{k^3 T}~~~~~~~~~~
\rht = \frac{\rho_3}{k}=\frac{\rho}{kT}~~~~~~~~~~
{\et}^2(k,\rho) = \frac{e^2_3  (k,\rho)}{k}=\frac{e^2(k,\rho,T)T}{k}.
\label{two} \ee
The evolution equations are \cite{gauge,abelian1,me}
\beq
\frac{\partial u_k}{\partial t} = & 
~-3 u_k + \rht u'_k
-v_3 L^3_0 \left( u'_k +2 u''_k \rht \right)
- 3 v_3 L^3_0 \left( u'_k \right)
- 6 v_3 L^3_0 \left( 2 \et^2 \rht \right),
\label{three} \\
\frac{\partial \et^2}{\partial t} = &
-\et^2 - \frac{4}{3} v_3~88~\et^4 l^3_{NA}
\theta \left( 2 \et^2 \rht \right).
\label{four} \eeq
Here $t=\ln(k/\Lx)$, 
$v_3=1/8 \pi^2$, $l^3_{NA}=0.677$
and primes denote derivatives with respect to $\rht$.
The ``threshold'' functions $L^3_0(w)$ take negative values, 
and fall off for large rescaled masses of
the radial scalar, Goldstone and gauge modes that serve as their 
arguments. In the evolution equation for the gauge coupling
the ``threshold'' function has been approximated by a $\theta$-function.
The various subscripts and superscripts 
set equal to 3 in the above expression indicate
that the evolution is typical of a three-dimensional theory. 
We started with the evolution equation for the
four-dimensional theory at non-zero temperature, and, 
for $k \ll T$, 
obtained the evolution equation
for an effective three-dimensional theory at zero temperature.
Thus we have obtained a realization of the mechanism of dimensional
reduction. In ref. \cite{me} 
the above equations were shown to be a very good approximation
in the whole region
$k \leq T$, for the parameter range of interest for the 
electroweak phase transition.
The renormalized theory (for $k=0$) at non-zero temperature is determined
through their numerical integration with algorithms discussed
in ref. \cite{num}.
The initial conditions at $k=T$ can be expressed in terms of
the renormalized parameters of the zero-temperature theory at $k=0$
(minimum of the potential $\rhz$, quartic coupling $\lx_R$,
and gauge coupling $e^2_R$, or, 
equivalently, gauge and Higgs boson masses, and gauge coupling) 
and the temperature. 

In fig. 1 we present the evolution 
of the effective average potential $U_k(\rho,T)$ for the 
$SU(2)$ Higgs model with 
$e^2_R=0.1073$,
$m_W=80.6$ GeV,
$m_H=35$ GeV,
as the 
coarse-graining scale $k$ is lowered.
The temperature is chosen near the critical one. 
Initially the potential has only one minimum 
away from the origin, which receives 
a renormalization proportional to the running scale $k$
during the evolution. 
At some point a new, shallow minimum appears at the origin. 
It is induced by 
the integration of thermal fluctuations, through the
generalization of the Coleman-Weinberg mechanism \cite{colwein}.
The evolution slows down at later stages and the 
potential converges towards a non-convex 
profile.  With an appropriate choice of 
the coarse-graining scale $k$, this potential 
can be employed for the proper discussion
of the properties of the first-order 
phase transition \cite{me,coarse}. 
During the same ``time'' $t$, the running gauge coupling 
$e^2(k,\rho,T)$ increases, because
the $SU(2)$ Higgs model is asymptotically free.
This evolution is stopped, however, because of 
the decoupling of the massive gauge
field fluctuations when the running scale $k$ becomes 
smaller than their mass $\sqrt{2 e^2 \rho}$. 
This decoupling takes place earlier for larger $\rho$, while
for small values of $\rho$ near the origin 
the gauge coupling continues to increase.
Eventually it reaches a critical value, 
for which a confining regime is expected to emerge.
An estimate for this value is $\alpha=4\et^2/4 \pi=1$.
(The reason for the factor of 4 is our unconventional
normalization of the gauge coupling.)
We have stopped the evolution at this
scale, which is determined through the expression
\footnote{
The value of 
$k_{conf}$ is not very sensitive on its precise 
definition. The gauge coupling $\et^2(k)$ becomes of order 1
at scales $k$ within 10 \% from the scale $k_{div}$ at which it diverges. 
}
\be
\et^2(k_{conf},\rht=0) = \frac{e^2(k_{conf},\rho=0,T)~T}{k_{conf}} = \pi.
\label{five} \ee
The evolution for $k< k_{conf}$ cannot be reliably described in terms 
of a parametrization of the 
effective average action through fundamental fields. 
Instead, one has to use a basis of
composite operators, corresponding to the bound states that 
are expected to form at these scales. The two formulations
must be matched at $k= k_{conf}$ \cite{bound}. 
The various bound states are expected to be massive, with 
masses of the order of $k_{conf}$, and they should soon decouple. 
The most significant contribution from this last part of the evolution
comes from the appearance of condensates associated with operators
such as 
$F^2$. In ref. \cite{condensate} it was shown that $F^2$ develops a
non-zero expectation value in the confining regime, 
with a reduction of the energy density compared to the 
state with $F^2=0$. This negative contribution affects the potential
of fig. 1. The reason is that the strongly coupled regime 
appears only near the origin of the potential, where the running
of the gauge coupling has not been cut off.
We expect, therefore, 
a negative contribution to the potential for the region around
$\rho=0$ where the coupling satisfies eq. (\ref{five}).
On dimensional grounds, 
we can approximate the negative contribution to the potential by 
\be
\Delta U = - \left( c k_{conf} \right)^3 T.
\label{six} \ee
The phenomenological constant $c$ is expected to be of order 1. We consider
the values $c=1$ and 0.5 in order to study the effect of this
parameter on the characteristics of the  phase transition. 
In fig. 1 we display the final form of the potential, if the 
above contribution with $c=1$ is added to the result of the integration
at $k=k_{conf}$.
This term is added to the potential only 
in the small region around the origin where the running 
gauge coupling 
reaches the value determined by eq. (\ref{five}).
As a result, the shallow minimum at the origin becomes as deep as the
minimum at non-zero $\rho$. 
We have not attempted to account for the $\rho$ dependence of the 
$F^2$ condensate beyond a crude step-like 
behaviour, by introducing additional phenomenological 
parameters. As our analysis
cannot provide any hint on the value of these parameters,
we have preferred to neglect them completely. 
This explains the steep rise of the potential of fig. 1 near the origin.
Notice however that the influence of this crude approximation 
on the characteristics of the phase transition is rather small.
For example, the latent heat is determined by the temperature 
dependence of the energy density at the minimum 
away from the origin. This depends on the value of the
contribution of eq. (\ref{six}), but not on 
its precise $\rho$ dependence. Similarly, the  
surface tension involves an 
integration over the whole range between the two minima, which reduces
the effect of our approximation. 

In fig. 2 we present the evolution of the
potential for $m_H=70$ GeV and a temperature near the critical one.
We observe the running of the minimum and 
the curvature at the origin becoming less negative. 
In the region around the origin, the running gauge coupling
reaches the critical value of eq. (\ref{five}), 
at which the confining regime is 
expected to set in. The size of this region relative to the 
location of the minimum of the potential is larger than in the case of 
fig. 1. 
Confinement sets in while the curvature at the origin
is still negative and the Coleman-Weinberg
mechanism has not yet become effective. 
At this point we stop the evolution and 
add the contribution of eq. (\ref{six}) 
to the potential (here we have used 
$c=0.5$), in the region around the origin where the coupling
has the critical value. 
This results in a new minimum between the origin 
and the original minimum of the potential. 
A first-order phase transition is predicted,
but both minima 
now exist at non-zero values of the Higgs field. 
Within the 
effective average action approach, 
the appearance of a minimum at a non-zero Higgs field expectation value has 
been observed for the first time in ref. \cite{bastian}, where a simplified 
version of the evolution equation for the potential has been used.
Within the gap-equation approach, it has been observed 
in ref. \cite{owe}. 
For even larger scalar masses a qualitative change 
is expected. The evolution resembles that in fig. 2, until the 
gauge coupling reaches its critical value. 
When the strongly coupled regime sets in, its extent is larger
than the location of the minimum of the potential. 
This indicates that no minimum exists any longer where the 
description in terms of fundamental fields is possible. 
The two-minimum structure of the potential is not expected to 
appear any more.
Instead, the transition to a 
non-perturbative vacuum is expected to 
be continuous, without appearance of singularities. 
As a result no phase transition exists any more. This indicates
the change from first-order phase transitions
to analytical crossovers for large scalar masses. Our analysis
predicts
a critical scalar mass in the range 80--100 GeV. 
The possibility of a crossover for gauge theories has been 
first suggested in ref. \cite{banks}.
For the electroweak phase transition, 
it has been observed in ref. \cite{owe}, 
where the gap-equation approach has been followed.  
Within the effective average action approach, it 
has been observed
in refs. \cite{runngauge,bastian}, where arguments
similar to ours have 
been given for the first time. The most reliable and detailed
results that confirm this
possibility have been obtained through the lattice approach \cite{crossover}.

The characteristics of the phase transition can be deduced from
the form of the potential, as explained in ref. \cite{me}.
In table 1 we present them for models with 
$m_W=80.6$ GeV, and $e^2_R$, $m_H$ equal to the set of values
employed in the lattice studies of refs. \cite{fodor1}--\cite{laine}.
The value $c=1$ has been used in the contribution of eq. (\ref{six}).
We give the critical temperature, the location of the 
minimum away from the origin, the location of the
confining minimum (whenever it appears), the scale $k_{conf}$
where confinement sets in,
the masses of the fields at the minimum away from the origin, 
the surface tension and
the latent heat of the phase transition.
All the above quantities have been extracted from the 
potential at $k=k_{conf}$ after the addition of the contribution
of eq. (\ref{six}). The surface tension has been calculated 
through the leading semiclassical thermal-tunnelling fluctuation \cite{me}.
We do not give any values for the masses of the fields at the origin. 
There, the proper description should include bound states instead
of fundamental fields; this is the subject of future work.

In table 2 we compare the renormalization-group results
with those of lattice and
perturbative studies for $e^2_R=0.1463$, $m_W=80.6$ GeV, 
$m_H=34$ GeV. We consider the critical temperature, surface tension
and latent heat that are discussed in refs. \cite{fodor3,fodor4,laine}. 
The discontinuity in the Higgs field expectation value in table 1
corresponds to 
expectation values
of $\phi$ in a formalism with gauge fixing. This is not 
comparable to the discontinuity of the 
expectation value of the operator $\phi^{\dagger} \phi$ that is calculated
in the lattice studies. 
The first column (RG1) results from the 
integration of the evolution equation for the potential, with 
the contribution of the strongly coupled regime around the origin
approximated by eq. (\ref{six}) with $c=1$. For the second column 
(RG2) we have used $c=0.5$. The comparison of the two columns
gives an estimate of the effect of the confining regime on the
properties of the phase transition. It is apparent from figs. 1 and 2
that the addition of the contribution of eq. (\ref{six}) results
in a more strongly first-order phase transition. 
This effect is enhanced for larger $c$, and 
is reflected in the larger values of 
$\sigma/\tcr^{3}$ and 
$\Delta Q/\tcr^{4}$. 
In the  third column (L1) we give the results of the lattice study of 
refs. \cite{fodor3,fodor4}. The continuum
limit $L_t=\infty$ has been studied only for $T_{cr}$.
For the surface tension and latent heat we list the
finite-lattice results. 
In the fourth column (L2) we list values that have 
been obtained in ref. \cite{laine} through the 
extrapolation of the results of ref. \cite{kaj1}
to the value $m_H=34$ GeV. These results have been
obtained through numerical simulations of the effective 
three-dimensional theory, so that the continuum limit for
the time direction is not needed.
Finally in the last two columns we give the perturbative 
predictions to order $g^3$ (PT1) and $g^4$ (PT2) \cite{fodor4}.
We observe that the RG1, RG2 predictions for the critical temperature 
are larger than the L1, L2 results, but smaller than 
the PT1, PT2 predictions. Moreover, the RG1 prediction
is significantly closer to the lattice results. This indicates 
that the value $c=1$ is the preferred one. This choice is further
supported by the comparison of the results for the surface
tension and latent heat. 
For $c=1$, the non-perturbative effects dramatically increase the 
strength of the phase transition. Their influence is much larger
than that due to the growth of the running gauge coupling.
This effect has been estimated
in ref. \cite{me}, where the form of the potential at the critical temperature
has been calculated, without and with the running of the gauge coupling
(by considering a zero $\beta$-function for the gauge coupling, or by
employing eq. (\ref{four})).  
As the first-order phase
transition is triggered by the gauge-field fluctuations, its
strength is increased when the running of the gauge coupling is taken into
account. An increase in the strength of the phase transition is
observed in higher orders of perturbation theory (compare columns
PT1 and PT2). However, the agreement of column PT2 with the lattice results
cannot be explained by our analysis. Our fig.
1 indicates that the non-perturbative effects of the
strongly-coupled regime 
are substantial already at $m_H=35$ GeV.
In summary, we conclude that, for scalar masses around 
35 GeV,  the characteristics of the first-order
phase transition are properly reproduced by our analysis,
if the non-perturbative contribution from the 
strongly-coupled regime in the evolution of the potential
is approximated by eq. (\ref{six}) with $c=1$. 
We should point out, however, that the determination of the value
$c=1$ through an explicit calculation in the context of the
effective average action is pending.

In table 3 we compare the renormalization-group approach with the
lattice studies for several scalar masses. We observe 
that the choice $c=1$ in eq. (\ref{six}) leads to a good description of
the phase transition for light Higgs bosons with $m_H=29.1,$ 34 GeV.
For larger $m_H$ the agreement between renormalization-group and 
lattice results becomes much weaker. The discrepancy is most obvious
for the surface tension. This quantity is the most difficult 
to extract from the renormalization-group calculation for 
large $m_H$, when the phase transition becomes very weakly first order.
This is apparent in fig. 2, where there is no interval in the evolution
during which the potential converges towards a stable non-convex profile
(in contrast to fig. 1). The determination of the surface tension from 
the potential, through a semiclassical tunnelling solution, is
very sensitive to the choice of coarse-graining scale. This
indicates that the contribution to the tunnelling rate 
coming from fluctuations around the 
semiclassical solution is important and must be 
taken into account \cite{coarse}. 
Also the step-like behaviour of the potential of fig. 2 becomes
a very crude approximation. 
As a result, the values for
the surface tension, obtained through the renormalization-group approach
for $m_H=49$, 54.4 GeV, are subject to large uncertainties.  
It should also be pointed out that the sensitivity of the
surface tension and latent heat on the choice of $c$ in 
eq. (\ref{six}) is large for $m_H=49$, 54.4 GeV. 
A value $c=0.5$ leads to the necessary reduction of 
$\sigma/\tcr^{3}$ and $\Delta Q/\tcr^{4}$. However, it also leads 
to a much larger value of $\tcr$ than the choice $c=1$.

In fig. 3 we plot the location of the two minima of the
potential at the critical temperature, as a function 
of the scalar mass, for $m_W=80.6$ GeV. 
The data of table 1 have been used, as well 
as those of ref. \cite{me}. An additional calculation has provided the
points at $m_H=40$ GeV. 
The stars correspond to $e^2_R \simeq 0.145$ and the 
boxes to $e^2_R \simeq 0.107$. The white boxes for $m_H=70$ GeV
are only indicative, as the quantitative reliability of the 
results is small for $m_H \gta 55$ GeV. An extrapolation of the 
two lines of data for the two minima of the potential indicates 
that they should coincide for a Higgs mass in the range 
80--100 GeV. This is the range in which an analytical crossover
is expected to set in. However,  a detailed description of the system in 
this range requires the parametrization of the 
effective average action in terms of composite operators,
corresponding to the bound states that are expected to appear.

In conclusion, the comparison of renormalization-group and
lattice studies reveals the mechanisms underlying the phenomenology
of the electroweak phase transition. 
The Coleman-Weinberg mechanism operates for scalar masses
$m_H \lta 40$ GeV. It generates a two-minimum structure in the 
potential and induces the first-order phase transition.
The strength of the phase transition is enhanced by the 
confining dynamics near the symmetric phase of the model, which 
increases the depth of the minimum of the potential at the 
origin by an amount given by eq. (\ref{six}). The 
choice $c=1$ for the phenomenological parameter in this equation
leads to agreement of the renormalization-group predictions 
with the lattice results for the properties 
of the phase transition. 
For $m_H \gta 40$ GeV the confining regime sets in above the 
energy scales at which 
the Coleman-Weinberg mechanism becomes effective. 
A two-minimum structure is generated, but both minima exist
for non-zero values of the Higgs field. 
The phase transition becomes progressively more weakly first order
as the scalar mass is increased. Finally, for 
a critical scalar mass in the range 80--100 GeV, the
two-minimum structure of the potential is no longer expected to appear.
The only minimum of the potential exists within the confining
region near the symmetric phase. 
An analytical crossover is expected to replace the 
first-order phase transition for 
larger scalar masses.

\noindent
{\bf Acknowledgements:}
We would like to thank Z. Fodor, M. Shaposhnikov and C. Wetterich
for helpful discussions.

\vspace{3cm}

\setcounter{equation}{0}
\renewcommand{\theequation}{{\bf F.}\arabic{equation}}

\section*{Figures}

\renewcommand{\labelenumi}{Fig. \arabic{enumi}}
\begin{enumerate}

\item  
The effective average potential $U_k(\rho,T_{cr})$, for the 
$SU(2)$ Higgs model with 
$e^2_R=0.1073$,
$m_W=80.6$ GeV,
$m_H=35$ GeV,
as the 
coarse-graining scale $k$ is lowered.
\vspace{5mm}

\item  
The effective average potential $U_k(\rho,T_{cr})$, for the 
$SU(2)$ Higgs model with 
$e^2_R=0.1073$,
$m_W=80.6$ GeV,
$m_H=70$ GeV, as the 
coarse-graining scale $k$ is lowered.
\vspace{5mm}

\item  
The location of the two minima of the potential at the critical temperature,
as a function of the Higgs boson mass, for $m_W=80.6$ GeV. 
The boxes correspond to $e^2_R \simeq 0.107$ and the 
stars to $e^2_R \simeq 0.145$.

\end{enumerate}

\newpage

\section*{Tables}

\begin{table} [h]
\renewcommand{\arraystretch}{1.5}
\hspace*{\fill}
\begin{tabular}{|c||c|c|c|c|}     
\hline
$e^2_R$ & 0.1463 & 0.1446 & 0.1073 & 0.1073
\\ \hline
$m_H$/GeV & 34 & 49 & 29.1 & 54.4
\\ \hline
$\tcr$/GeV 
& 77.5 & 103 & 78.7 & 130
\\ \hline 
$\tcr/m_H$
& 2.28 & 2.10 & 2.71 & 2.39
\\ \hline 
$\phi_0(\tcr)$/GeV 
& 110 & 114 & 118 & 119
\\ \hline 
$\phi_{np}(\tcr)$/GeV 
& --- & 31.2 & --- & 31.7 
\\ \hline 
$\Delta \phi(\tcr)/\tcr$ 
& 1.42 & 0.804 & 1.50 & 0.672
\\ \hline 
$k_{conf}$/GeV 
& 13.0 & 17.3  & 10.1 & 16.6
\\ \hline 
$m_W \left( \rhz,\tcr \right)$/GeV 
& 44.9 & 47.1 & 41.3 & 43.9
\\ \hline 
$m_H \left( \rhz,\tcr \right)$/GeV
& 13.8 & 21.8 & 10.1 & 21.6
\\ \hline 
$\sigma$/(GeV$)^3$ 
& $3.04 \times 10^{4}$ & $ 4.75 \times 10^{4}$
& $2.63 \times 10^{4}$ & $ 5.36 \times 10^{4}$
\\ \hline 
$\sigma/\tcr^{3}$ 
& $6.53 \times 10^{-2}$ & $ 4.35 \times 10^{-2}$
& $5.40 \times 10^{-2}$ & $ 2.44 \times 10^{-2}$
\\ \hline 
$\Delta Q$/(GeV$)^4$ 
& $6.84 \times 10^{6}$ &  $ 1.63 \times 10^{7}$  
& $6.07 \times 10^{6}$ & $  1.86 \times 10^{7}$ 
\\ \hline 
$\Delta Q/\tcr^{4}$ 
& $1.90 \times 10^{-1}$ & $ 1.45 \times 10^{-1}$ 
& $1.58 \times 10^{-1}$ & $ 6.51 \times 10^{-2}$ 
\\ \hline 
\end{tabular}
\hspace*{\fill}
\renewcommand{\arraystretch}{1}
\caption[y]
{
Characteristics of the first-order phase transition 
for the $SU(2)$ Higgs model with $m_W=80.6$ GeV.} 
\end{table}

\begin{table} [ht]
\renewcommand{\arraystretch}{1.5}
\hspace*{\fill}
\begin{tabular}{|c||c|c|c|c|c|c|}     
\hline 
& RG1 & RG2 & L1 & L2 & PT1 & PT2 
\\ \hline \hline 
$\tcr$/GeV 
& 77.5 & 79.6 
& 73.0(14) & 75.7(5) 
& 78.2 & 78.0
\\ \hline 
$\sigma/\tcr^{3}$ 
& 0.0653 & 0.0274
& $0.053(5)~~(L_t=2)$ & 0.050(12) 
& 0.027 & 0.056
\\ \hline 
$\Delta Q/\tcr^{4}$ 
& 0.190 & 0.120
& $0.28(12)~~(L_t=4)$ & 0.19(1)
& 0.13 & 0.19
\\ \hline 
\end{tabular}
\hspace*{\fill}
\renewcommand{\arraystretch}{1}
\caption[y]
{
Comparison of results from renormalization-group, lattice and
perturbative studies for $e^2_R=0.1463$, $m_W=80.6$ GeV, 
$m_H=34$ GeV. \\
RG1: Integration of the evolution equation for the potential, with 
the contribution of the strongly coupled regime around the origin
approximated by eq. (\ref{six}) with $c=1$. \\
RG2: The same as for RG1, but with $c=0.5$. \\
L1: Lattice study of refs. \cite{fodor3,fodor4}. The continuum
limit $L_t=\infty$ has been studied only for $T_{cr}$. \\
L2: Extrapolated results \cite{laine} of the lattice 
study of ref. \cite{kaj1}. \\
PT1: Perturbation theory to order $e^3$ \cite{fodor4}.\\
PT2: Perturbation theory to order $e^4$ \cite{fodor4}. 
}
\end{table}

\begin{table} [ht]
\renewcommand{\arraystretch}{1.5}
\hspace*{\fill}
\begin{tabular}{|c||c|c||c|c||c|c||c|c|}     
\hline 
& RG & L1 & RG & L2 & RG & L3 & RG & L3
\\ \hline \hline
$e^2_R$
&\multicolumn{2}{c||}{0.1463} 
&\multicolumn{2}{c||}{0.1446} 
&\multicolumn{2}{c||}{0.1073} 
&\multicolumn{2}{c|}{0.1073} 
\\ \hline 
$m_H$/GeV 
&\multicolumn{2}{c||}{34} 
&\multicolumn{2}{c||}{49} 
&\multicolumn{2}{c||}{29.1} 
&\multicolumn{2}{c|}{54.4} 
\\ \hline 
$\tcr$/GeV 
& 77.5 & 73.0(14) 
& 103 & [92.9(13)]
& 78.7 & 76.8(5)
& 130 & 132.6(5)
\\ \hline 
$\sigma/\tcr^{3}$ 
& 0.0653 & [0.053(5)]
& 0.0435 & [0.008(2)] 
& 0.0540 & [0.071(3)]
& 0.0244 & 0.0017(4)
\\ \hline 
$\Delta Q/\tcr^{4}$ 
& 0.190 & [0.28(12)] 
& 0.145 & [0.122(9)]
& 0.158 & 0.200(7)
& 0.0651 & 0.0294(7)
\\ \hline 
\end{tabular}
\hspace*{\fill}
\renewcommand{\arraystretch}{1}
\caption[y]
{
Comparison of results from renormalization-group and lattice studies 
for $m_W=80.6$ GeV. \\
RG: Integration of the evolution equation for the potential. \\
L1: Lattice study of refs. \cite{fodor3,fodor4}. \\
L2: Lattice study of ref. \cite{fodor1}. The results correspond
to values of 
$e^2_R$ slightly different from the one quoted in the first row. \\
L3: Lattice study of ref. \cite{kaj1}.
 \\
For the results in brackets, the continuum limit has
not been studied.
} 
\end{table}

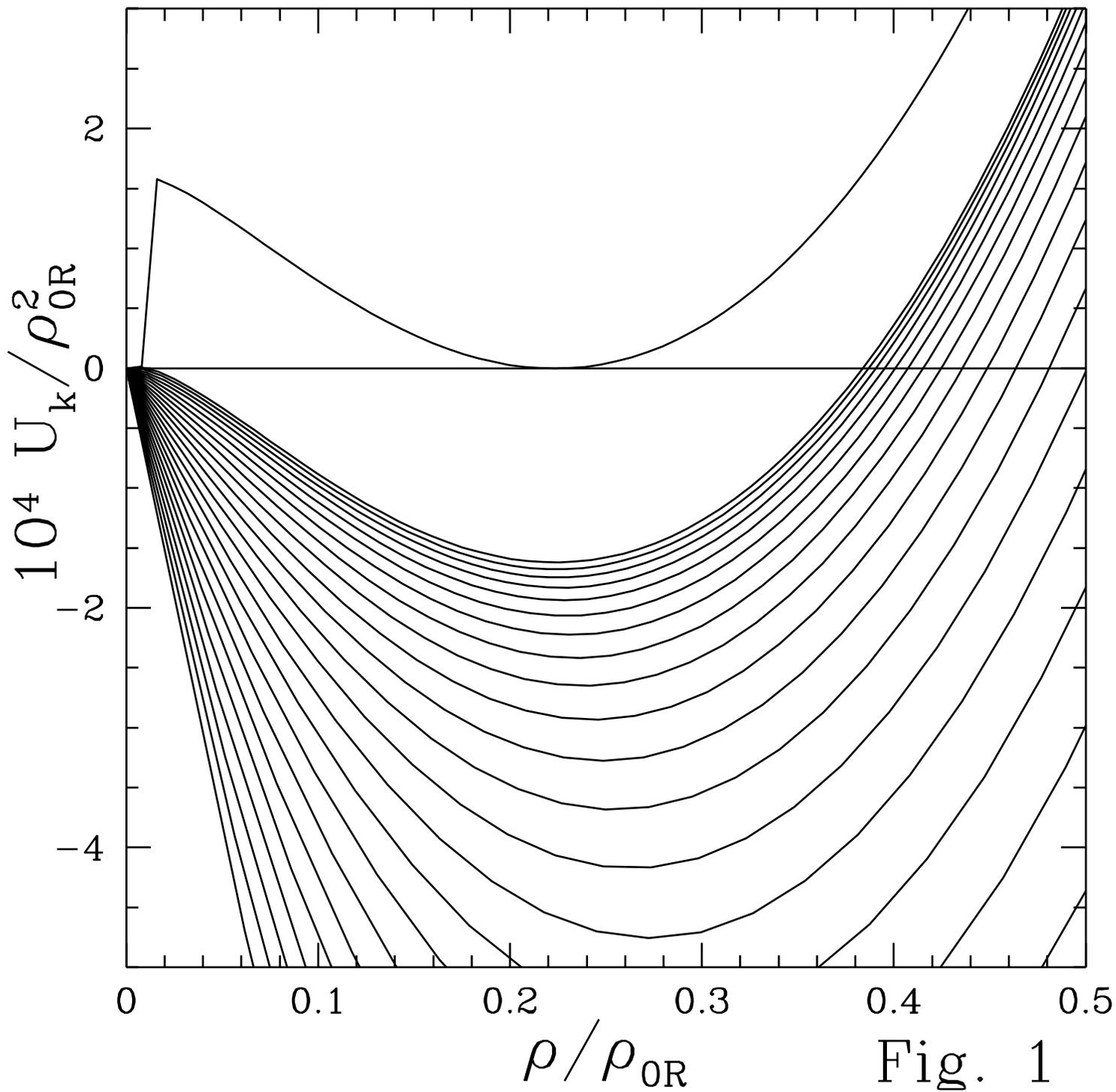
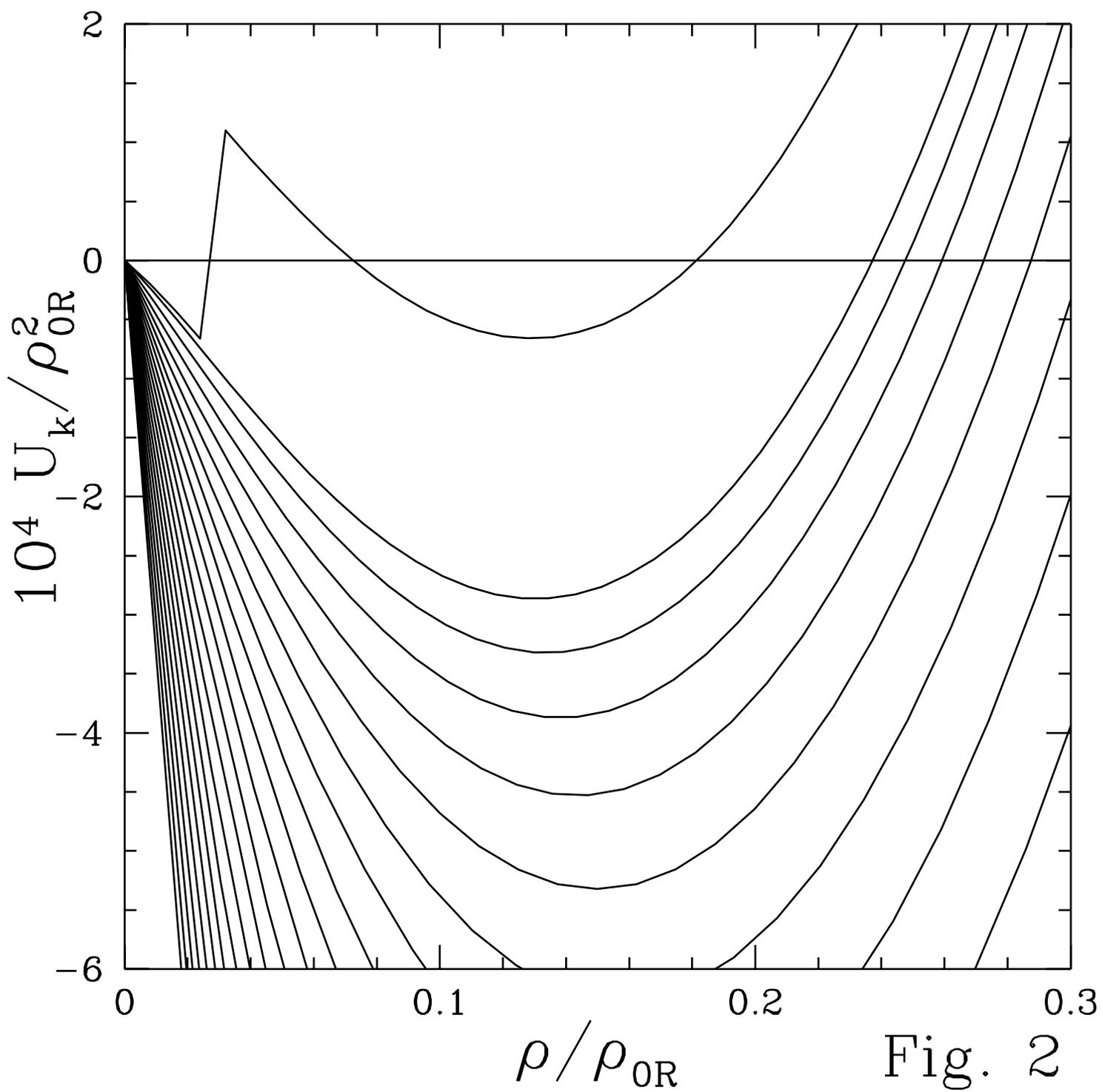
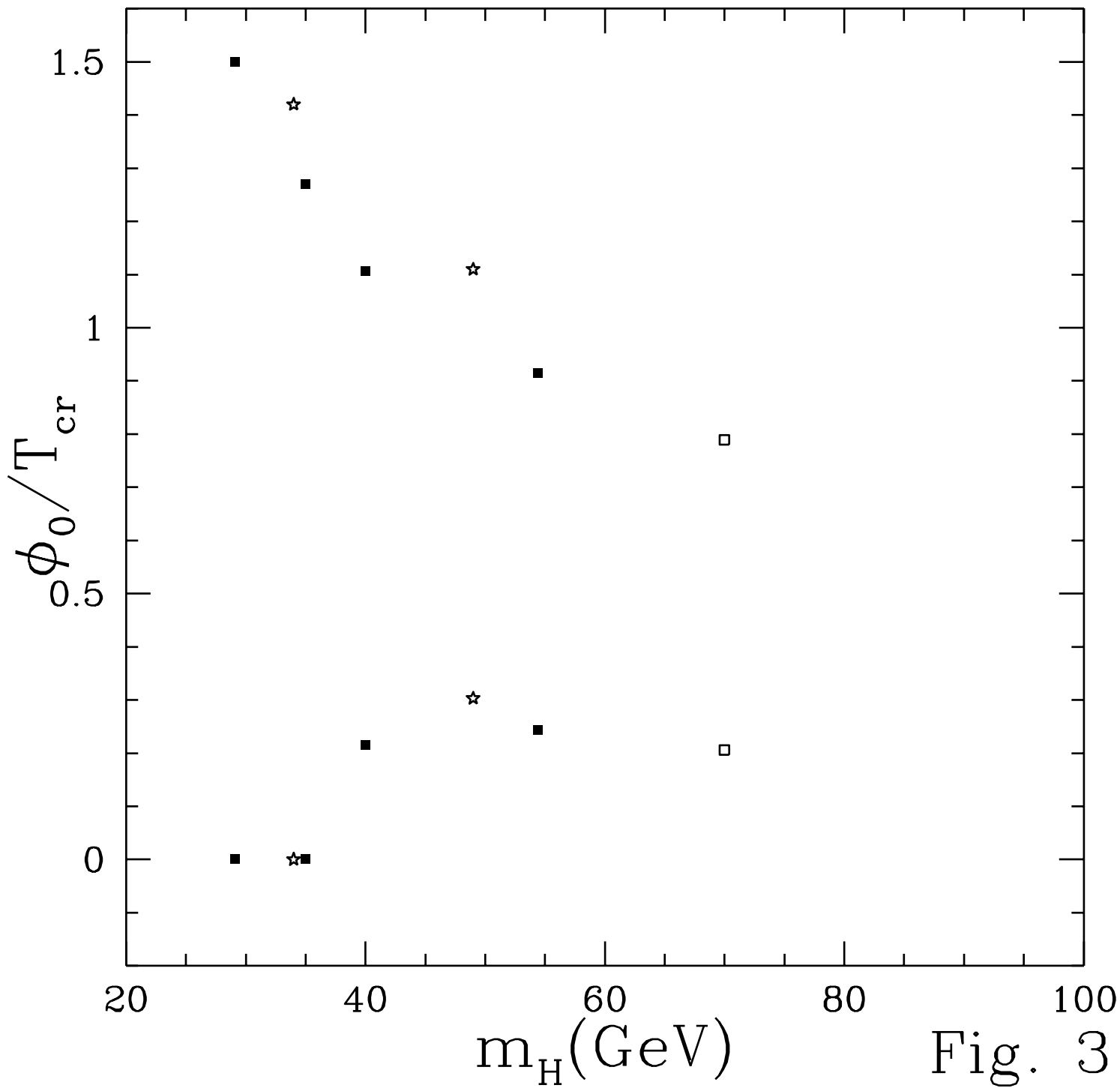
\end{document}